\documentclass[12pt,preprint]{aastex}

\begin{document}

\title{Evolution of the Solar Nebula. VIII. Spatial and Temporal
Heterogeneity of Short-Lived Radioisotopes and Stable Oxygen Isotopes}

\author{Alan P.~Boss}
\affil{Department of Terrestrial Magnetism, Carnegie Institution of 
Washington, 5241 Broad Branch Road, NW, Washington, DC 20015-1305}
\authoremail{boss@dtm.ciw.edu}

\begin{abstract}

Isotopic abundances of short-lived radionuclides such as $^{26}$Al
provide the most precise chronometers of events in the early solar 
system, provided that they were initially homogeneously distributed.
On the other hand, the abundances of the three stable isotopes of 
oxygen in primitive meteorites show a mass-independent fractionation 
that survived homogenization in the solar nebula. As as result of 
this and other cosmochemical evidence, the degree of spatial 
heterogeneity of isotopes in the solar nebula has long been a puzzle. 
We show here that based on hydrodynamical models of the mixing and 
transport of isotopic anomalies formed at, or injected onto, the 
surface of the solar nebula, initially high levels of isotopic spatial 
heterogeneity are expected to fall to steady state levels ($\sim$ 10\%) 
low enough to validate the use of $^{26}$Al for chronometry, but high
enough to preserve the evidence for mass-independent fractionation 
of oxygen isotopes. The solution to this puzzle relies on the mixing being
accomplished by the chaotic fluid motions in a marginally gravitationally
unstable disk, as seems to be required for the formation of gas giant
planets and by the inability of alternative physical processes to drive 
large-scale mixing and transport in the planet-forming midplane of 
the solar nebula. Such a disk is also capable of large-scale outward 
transport of the thermally annealed dust grains found in comets, 
and of driving the shock fronts that appear to be responsible for 
much of the thermal processing of the components of primitive meteorites,
creating a self-consistent picture of the basic physical processes shaping 
the early solar nebula.

\end{abstract}

\keywords{accretion, accretion disks -- hydrodynamics -- solar system:
formation -- planetary systems}

\section{Introduction}

 Isotopic evidence has been presented for both homogeneity 
(e.g., Hsu, Huss \& Wasserburg 2003; Guan et al. 2006; 
Thrane et al. 2006) and heterogeneity (e.g., Simon et al. 1998;
Sugiura, Miyazaki, \& Yin 2006) of short-lived radioisotopes 
such as $^{26}$Al and $^{60}$Fe in the the solar nebula. A high 
degree of homogeneity is required if the inferred starting ratios of
$^{26}$Al/$^{27}$Al are to be used as precise chronometers for the early 
solar system (Bizarro et al. 2004; Halliday 2004; Krot et al. 2005b). 
The degree of uniformity is equally murky even for stable isotopes 
such as those of molybdenum, for which both heterogeneity 
(Yin, Jacobsen \& Yamashita 2002; Dauphas, Marty \& Reisberg 2002)
and homogeneity (Becker \& Walker 2003) have been asserted.
For samarium and neodymium isotopes, both nebular homogeneity and 
heterogeneity have been claimed, depending on the stellar 
nucleosynthesis source (Andreasen \& Sharma 2006). 
The three stable isotopes of oxygen, however, show clear
evidence for heterogeneity in primitive meteorites (Clayton 1993).
At some level, isotopic heterogeneity must exist. The question 
then becomes, what level of heterogeneity is to be expected?

 The short-lived radioisotope (SLRI) $^{60}$Fe must have been 
synthesized in a supernova (Mostefaoui, Lugmair \& Hoppe 2005;
Tachibana et al. 2006) and then injected into the presolar cloud 
(Vanhala \& Boss 2000, 2002) or onto the surface of the solar nebula 
(Desch \& Ouellette 2005; Ouellette \& Desch 2006).
A similar nucleosynthetic event is likely to be the source of the 
bulk of the solar nebula's $^{26}$Al (Limongi \& Chieffi 2006;
Sahijpal \& Soni 2006; however, see Gounelle et al. 2006 for a 
contrary point of view, and Bizzaro et al. 2006 for evidence that the 
$^{60}$Fe was injected after the $^{26}$Al). This injection occurred 
through narrow Rayleigh-Taylor (R-T) fingers (Vanhala \& Boss 2000, 2002) 
that peppered the disk surface with highly nonuniform doses of 
SLRIs. The scatter in initial $^{26}$Al/$^{27}$Al ratios observed in
meteoritical components ranging from values of $\sim 0$ to 
$\sim 4.5 \times 10^{-5}$ (MacPherson, Davis \& Zinner 1995)
or even higher values $\sim 7 \times 10^{-5}$ (Young et al. 2005)
can be attributed to spatial heterogeneity, temporal heterogeneity
(i.e., crystallization at different times, allowing for decay of the
SLRIs), or some combination of these two extremes. The fact that these
SLRIs were injected into the solar nebula means that immediately following
the arrival of each R-T finger, the nebula must have been strongly spatially
heterogeneous in terms of the abundances of those isotopes, possibly
explaining some of the observed range of $^{26}$Al/$^{27}$Al ratios. 

 The wide range in stable oxygen isotope abundances
is best explained by self-shielding of molecular CO gas from 
UV photodissociation at the surface of the solar nebula (Clayton 2002;
Lyons \& Young 2005; however, see Marcus 2004 for an opposing view) 
or in the presolar molecular cloud (Yurimoto \& Kuramoto 2004)
and is not associated with R-T injection events. Oxygen isotopic anomalies 
formed at the disk surface most likely could only have been preserved 
in the outer nebula, however, so the oxygen isotope anomalies may have 
also originated in a spatially heterogeneous manner. Furthermore,
dust-gas fractionation is required in order to affect stable oxygen
isotope ratios, and this fractionation process would have been a function 
of both time and space in the disk.

 Besides this evidence for initial spatial heterogeneity, there is 
also strong evidence for large-scale transport of isotopes and small 
grains. The presence of both high- and low-temperature phases of silicates
in comets requires inward transport of amorphous silicates and
outward transport of annealed silicates over large distances
(several AU at least) in the solar nebula (Lisse et al. 2006).
Isotopic evidence suggests (Bizarro, Baker \& Haack 2004) that 
some Allende chondrules formed with an $^{26}$Al/$^{27}$Al ratio 
similar to that of calcium, aluminum-rich inclusions (CAIs). 
Chondrule formation thus appears to have begun shortly after CAI 
formation, and to have lasted for $\sim$ 1 to 4 Myr (Amelin et al. 2002). 
While chondrules are generally believed to have been melted by
shock-front heating at asteroidal distances (Iida et al. 2001; 
Desch \& Connolly 2002; Ciesla \& Hood 2002), CAIs are generally 
thought to have formed much closer to the protosun, because of 
their highly refractory compositions (Wood 2004). 
On the other hand, recent kinetic evaporation-condensation models 
(Alexander 2004) have shown that Type A (olivine-rich) chondrules,
compact Type A (melilite-rich) CAIs, and compact Type B (pyroxene-rich) 
CAIs may have formed at asteroidal distances, while Type B (pyroxene-rich) 
chondrules, Al-rich chondrules, and Type C (anorthite-rich) CAIs may have 
formed closer to the protosun. Either way, mixing of solids from the inner
solar nebula out to asteroidal distances seems to be required in order
to assemble chondrules, CAIs, and matrix grains into the chondritic 
meteorites (Boss \& Durisen 2005).

 Solids formed close to the protosun may have been carried upward by 
the protosun's bipolar outflow and lofted onto trajectories that would 
return them to the surface of the solar nebula at much greater distances 
(Shu et al. 2001). Alternatively, solids might have been transported outward 
to asteroidal or cometary distances by mixing processes within the solar 
nebula, such as generic turbulence or the mixing produced by spiral arms in 
a marginally gravitationally unstable nebula (Boss 2004a). In the Boss 
(2004a) models, however, the tracers representing solids tied to the gas were 
injected into the disk midplane or disk surface in rings or sectors
of rings centered on a distance of 9 AU, right in the middle of
the most gravitationally active region of the disk. These tracers
were then found to be rapidly (in less than 0.001 Myr) transported
both inward to 4 AU and outward to 20 AU, the inner and outer
boundaries of the computational grid, respectively. The question
then arises as to the fate of solids injected onto the
surface of the nebula at much greater distances, or residing inside
the disk closer to the protostar: can these solids be transported 
inward, or outward, respectively, through the chaotic region where the
gas giant planets are trying to form?

 We present here a new set of three dimensional hydrodynamical models
of mixing and transport in the solar nebula that attempts to answer
both of these questions about transport and mixing in protoplanetary
disks, with an emphasis on the solar nebula. These models are 
identical to those of Boss (2004a) except that in the new models the 
color fields representing isotopically distinct parcels of gas or
solids are introduced at radial distances of either 6 or 15 AU, 
instead of at the original 9 AU. 

\section{Numerical Methods}

 The disk evolution calculations were performed with a numerical code
that uses finite differences to solve the three dimensional equations 
of hydrodynamics, radiative transfer, and the Poisson equation for the
gravitational potential. The code is the same as that used in previous
studies of transport in disks (e.g., Boss 2004a, 2006) and has been shown to
be second-order-accurate in both space and time through convergence testing
(Boss \& Myhill 1992). The equations are solved on a spherical coordinate
grid. The number of grid points in each spatial direction is: $N_r = 101$,
$N_\theta = 23$ in $\pi/2 \ge \theta \ge 0$, and $N_\phi = 256$, for
a total of over one million grid points in both hemispheres. The radial 
grid is uniformly spaced between 4 and 20 AU, with boundary conditions 
at both the inner and outer edges chosen to absorb radial velocity
perturbations. The $\theta$ grid is compressed into the midplane to ensure
adequate vertical resolution ($\Delta \theta = 0.3^o$ at the midplane).
The $\phi$ grid is uniformly spaced, to prevent any bias in the azimuthal
direction. The central protostar wobbles in response to the growth of
nonaxisymmetry in the disk, thereby rigorously preserving the location of
the center of mass of the star and disk system. The number of terms in the
spherical harmonic expansion for the gravitational potential of the disk
is $N_{Ylm} = 32$. 

 As in Boss (2004a, 2006), the models treat radiative transfer in the
diffusion approximation, which should be valid near the disk midplane
and throughout most of the disk, because of the high vertical optical
depth. Artificial viscosity is not employed. The energy equation is solved
explicitly in conservation law form, as are the four other hydrodynamic
equations. 

\section{Turbulent Diffusion}
 
 Diffusion of the dust grains carrying the SLRI or oxygen isotope
anomalies with respect to the disk gas is handled through a modification
of the color equation (e.g., Foster \& Boss 1997; Boss 2004a). This
modification involves adding in to the right hand side of the color 
equation a diffusion term, i.e., second space derivative of the color 
field, multiplied by an appropriate diffusion coefficient $D$ (e.g., 
Stevenson 1990). For $D = constant$, the simplest assumption, the
equation of motion for the color field density $\rho_c$ is then

$${\partial \rho_c \over \partial t} + \nabla \cdot (\rho_c {\bf v}) =
D \nabla^2 \rho_c,$$

\noindent where ${\bf v}$ is the disk gas velocity and $t$ is the time. The 
color equation is solved in the same manner as the five other equations of 
motion, using finite differences and explicit time differencing.

 The models assume (Boss 2004a) that the eddy diffusivity $D$ can be 
approximated by the eddy viscosity of a classical viscous accretion 
disk, i.e., $D = \alpha h c_s$, 
where $\alpha =$ disk turbulent viscosity parameter, $h =$ disk
scale height, and $c_s =$ isothermal sound speed at the disk midplane.
The same assumption is made by Willacy et al. (2006) in their one
dimensional models of chemical effects in the outer regions of protoplanetary 
disks. For typical nebula values at 5 to 10 AU ($h \approx 1$ AU, $T 
\approx 100$ K, $c_s \approx 6 \times 10^4$ cm s$^{-1}$), the eddy
diffusivity becomes $D = 10^{18} \alpha$ cm$^2$ s$^{-1}$.

 While the color field is subject to eddy diffusivity when $\alpha \ne 0$, 
the underlying gaseous disk is not -- the gaseous disk remains effectively 
inviscid with an eddy viscosity $\nu = 0$. The eddy diffusivity is thus
assumed to be much larger than the eddy viscosity of the gas, i.e., 
$D >> \nu$, or $k = D/\nu >> 1$ (Stevenson 1990). Equivalently, this
assumes that the Schmidt number $Sc = \nu/D \approx 0$. For $k$ = 1 to 3, 
turbulent diffusion must work its way upstream against the inward flow 
onto the protosun, and Stevenson (1990) showed that little upstream 
mixing occurs in that case. However, Prinn (1990) argued that $k >> 1$ 
was likely to occur during early phases of disk evolution when 
the disk is still accreting material from the infalling cloud
envelope, which is the phase of evolution under investigation here.
Prinn (1990) pointed out that some turbulent eddies could transport 
angular momentum inward rather than outward, resulting in a negative 
contribution to the total disk viscosity, lowering the effective
value of $\nu$ and resulting in a large value of $k$. Disks with large 
$k$ are expected to be able to transport tracers upstream against
the inward flow onto the protosun.

 Whether $k >> 1$ or not would depend on the physical mechanisms
driving the disk turbulence $\nu$ and dust grain diffusion $D$.
For example, in a disk where turbulence is driven by vertical
convective motions (as in the present models; see Boss 2004b),
Stone et al. (2000) showed that convective cells are sheared by
differential rotation to such an extent that the net transport of
angular momentum is very small, i.e., $\nu \sim 0$. On the
other hand, the magnetorotational instability (MRI) in an ionized
disk has been shown to lead to rapid outward angular momentum
transport (Stone et al. 2000), so in an MRI-driven disk, one
might expect $k = 1/Sc \sim 1$, as has been found in three 
dimensional MRI models of mixing in the outer solar nebula by 
Turner et al. (2006). However, MRI effects are limited to
the disk's surfaces and inner regions (inside $\sim 1$ AU) by
the need for appreciable fractional ionization (Stone et al. 2000).
The midplanes of protoplanetary disks are essentially unionized in 
the planet-forming regions from $\sim 1$ to $\sim 15$ AU, preventing 
MRI from serving as a driver of midplane disk evolution in these 
regions (Matsumura \& Pudritz 2006). One is thus left with
the gravitational torques in a marginally gravitationally unstable
disk in order to drive global angular momementum transport.
Such torques are quite effective in driving disk evolution on
timescales as short as $\sim 10^5$ yrs (Boss 1998), while at
the same time transporting tracers both far upstream and downstream
(Boss 2004a). The latter models thus provide strong support for
the situation envisioned by Prinn (1990): $k >> 1$, because
gravitational torques and convective motions are able to 
transport tracers far upstream, even in the absence of viscous 
evolution of the disk ($\nu = 0$) and of additional diffusive transport 
of the tracers ($D \approx 0$). The present models continue in this
promising vein of investigation.

\section{Initial Conditions}

 The models consist of a $1 M_\odot$ central protostar surrounded
by a protoplanetary disk with a mass of 0.091 $M_\odot$ between 4 and
20 AU, as in Boss (2004a, 2006). Disks with similar masses appear to 
be necessary to form gas giant planets by core accretion (e.g.,
Inaba et al. 2003) or by disk instability (e.g., Boss et al. 2002).
The disks start with an outer disk temperature $T_o = 50$ K, leading
to a minimum in the Toomre $Q$ value of 1.5 from $\sim$ 8 to
$\sim$ 20 AU; inside 8 AU, $Q$ rises to a value of $\sim$ 9 because
of the much higher disk temperatures closer to the protosun.
A $Q$ value of 1.5 implies marginal instability to the growth
of gravitationally-driven perturbations, while $Q \sim 9$ implies
a high degree of stability.

 A color field representing SLRI or oxygen anomalies is sprayed onto 
the outer surface of an already evolving disk at a radial distance of 
either 6 or 15 AU, into a 90 degree (in the azimuthal direction) 
sector of a ring of width 1 AU, simulating the arrival of a R-T finger. 
The models include the effects of the diffusion of the color field with 
respect to the gaseous disk (described above) by a generic turbulent 
viscosity characterized by alpha disk parameters of $\alpha = 0$, 0.0001, 
or 0.01. The low $\alpha$ value (0.0001) is small enough to have only a 
minor effect on the color field (Boss 2004a). For $\alpha \le 0.0001$, the 
color field is transported and mixed primarily by a combination of the global 
actions of gravitational torques in the marginally gravitationally unstable 
disks and the local actions of convective-like motions (Boss 2004b).

 SLRI or oxygen anomalies that reside in the gas or in particles with sizes of 
mm to cm or smaller will remain tied to the gas over timescales of $\sim$ 
1000 yrs or so, because the relative motions caused by gas drag result in 
differential migration by distances of less than 0.1 AU in 1000 yrs, 
which is negligible compared to the distances they are transported 
by the gas in that time, justifying their representation by the
color field. In addition, solids in a disk with spiral arms will not 
migrate monotonically toward the protosun to be lost to melting 
(Cuzzi \& Zahnle 2004), but rather will be driven by gas drag forces to
the centers of the spiral arms, where they can survive indefinitely 
(Haghighipour \& Boss 2003a,b).

\section{Results}

 We present here the results of six models, representing injection
of the color field onto the disk surface at distances of either 6 AU
or 15 AU, for three values of the eddy diffusivity of the color
field with respect to the gas, quantified by $\alpha  = 0.01$ (H), 
0.0001 (L), or 0.0 (Z), for models 6H or 15H, 6L or 15L, and 6Z or 15Z,
respectively.

 The color fields shown in the figures represent the number density
of small solids (e.g., chondrules, CAIs, ice grains, or their precursor 
grain aggregates) in the disk carrying either SLRI or oxygen isotope
anomalies. In all the models, injection occurs after the disk has been
evolving for a time of 200 yr, sufficient time for a marginally
gravitationally unstable disk to develop a strongly non-axisymmetric
pattern of spiral arms and clumps (Boss 2004a). In all the models,  
downward (and upward) transport occurs on short timescales 
($\sim$ 30 yr) as a result of convective-like motions driven by the 
superadiabatic vertical temperature gradients between the disk's 
midplane and its upper layers that develop as the disk evolves
and forms dense spiral arms and clumps in the midplane (Boss 2004b).

\subsection{Inner Disk Injection}

 Model 6L involves a low level of color field diffusivity and
injection at 6 AU. Fig. 1 shows that the color field initially sprayed 
onto the disk's surface at a distance of 6 AU is transported down to 
the midplane of the disk in less than 30 yr. Once the 
color field reaches the midplane, the effect of the gravitationally
driven disk evolution is to transport the color field inward to 4 AU 
as well as outward to the disk boundary at 20 AU in less than 1000 yr 
(Fig. 2). For the next several 1000 yr, the color field continues to 
circulate around the disk and to mix with the disk gas (Fig. 3). Note
that the color field piles up artificially on the inner and outer
boundaries of the grid (Fig. 3); in a more realistic calculation,
these color fields would disappear from the numerical grid, as
the color was transported closer to the central protostar and
to the portions of the disk lying beyond 20 AU.

 Model 6Z, with no eddy diffusivity at all, was started from
model 6L at a time of 2506 yr. After almost another 1000 yr
of evolution with $\alpha = 0$, Fig. 4 shows that model 6Z
has evolved in very much the same manner as model 6L (cf., Fig. 3)
at the same time. This shows that in model 6L, the effects
of nonzero diffusivity are quite small in comparison to
the mixing and transport that is accomplished by the gravitational
torques and convective-like motions in the marginally 
gravitationally unstable disk.

 The color fields shown in Figs. 3 and 4 represent the number density
of SLRIs in disks after $\sim 3283$ yrs of evolution following
the injection event. Given the strong gradients in the color
fields evident in Figs. 3 and 4, it is clear that the color field
is highly spatially heterogeneous. The underlying gas density
distribution is equally highly non-axisymmetric, of course, and it
is the gas density that drives the transport of the color field.
What is really important to understand is the extent to which
the color field has become homogeneous with respect to the disk gas.
This is because cosmochemical SLRI abundances typically 
are given as ratios, i.e., $^{26}$Al/$^{27}$Al, where the SLRI
$^{26}$Al is derived from the injection event, whereas the reference stable
isotope $^{27}$Al is presumably derived primarily from a well-mixed presolar 
cloud. Hence in order to determine the variations in the SLRI ratio
$^{26}$Al/$^{27}$Al, the color field must be normalized
by the gas density. This has been done in Fig. 5, which
shows the log of the color/gas ratio for model 6L at the same time as
the color field in Fig. 3. It can be seen that the color/gas ratio
is remarkably uniform throughout the disk, with variations of
no more than 0.1 in the log, or no more than a factor of 1.26.
The only exceptions to this uniformity are regions close to
the inner and outer boundary where there is very little
color and even less gas density, so that the ratio becomes
large. Because these regions contain essentially no gas or color, 
these ratio variations are negligible. For comparison, Fig. 6
shows the color/gas ratio for model 6L at the same time as Fig. 1, 
soon after the color field has reached the midplane and before
it has had time to mix with the gas: at this time the color/gas
ratio field is strongly non-uniform wherever there is color.
Within a few 1000 yr, this strong initial heterogeneity is
largely erased by disk mixing and transport.

\subsection{Outer Disk Injection}

 While both the inner and outer disk injection models are appropriate
for considering SLRIs injected by R-T fingers, the outer disk 
injection models are considerably more appropriate for the mixing and
transport of oxygen isotope anomalies than the 6 AU models, because 
such anomalies are most likely to survive only when formed at the disk 
surface at distances of 15 AU and beyond (Lyons \& Young 2005).

 Fig. 7 shows model 15L at about the same time as model 6L in Fig. 2,
about 550 yr after the injection event. The color field has been
transported throughout more of the disk than in model 6L at this
time, as in model 15L the color field is able flow rapidly
downstream toward the protostar, whereas in model 6L, the color
field must fight its way upstream against this flow, and hence
the large-scale transport is slower than in model 15L.
The previous models (Boss 2004a), where injection incurred at 9 AU,
display color field evolutions intermediate between those of the 
models with injection at 6 and 15 AU, as expected.
 
 Fig. 8 shows the color/gas density ratio in the midplane of 
model 15L after 3491 yr, about the same time as that for
model 6L, shown in Fig. 5. Once again the color/gas ratio
has been transformed from a highly spatially heterogeneous
distribution into a more nearly uniform distribution, with
the exception of very low density regions close to the
inner and outer boundaries. 

\subsection{Evolution of the Dispersion}

 While Figs. 5 and 8 demonstrate the approach to spatial
homogeneity in these models, a more valuable quantity to
calculate is the dispersion of the color/gas density ratio
from its mean value, in order to make a more useful comparison with
distributions of isotopic measurements.

 Figs. 9 and 10 show the expected level of spatial heterogeneity
in SLRI ratios such as $^{26}$Al/$^{27}$Al as a function of time
in the solar nebula, following a single SLRI injection event, for
models 6L and 15L, respectively. The figures give the 
square root of the sum of the squares of the color field 
divided by the gas density subtracted from the mean value of the 
color field divided by the gas density, where the sum
is taken over the midplane grid points and is normalized by
the number of grid points being summed over. The sum excludes
the regions close to the inner and outer boundaries, as well
as regions with disk gas densities less than $10^{-12}$ g cm$^{-3}$,
as the low gas densities in these regions skew the calculations
of the dispersion of the ratio of color to disk gas, and these
regions contain comparatively little gas and dust. 

 Figs. 9 and 10 show the evolution of the dispersion (or standard deviation) 
of the SLRI ratio from the mean value. The injection transient dies away
on a timescale of $\sim$ 1000 yrs, as expected, and in both models seems to 
reach a steady state dispersion level of $\sim$ 10\%. This steady
state value is more clearly reached in model 15L than in 6L, because
of the longer time span covered by model 15L and because the colors
are transported primarily downstream in model 15L, allowing a faster
approach to a steady state, as is clear from comparing Figs. 9 and 10.
By way of comparison, at
the moment of injection, the ratio of color to gas density was highly 
spatially heterogeneous, being equal to zero everywhere except at the surface 
of the disk in an azimuthal sector at a distance from the protosun
of between 5.5 and 6.5 AU in model 6L or between 14.5 and 
15.5 AU in model 15L.

 In order to determine if the $\sim$ 10\% level of dispersion
is caused by the choice of the eddy diffusivity $D$, models 6L
and 15L were continued with the eddy diffusivity turned off,
i.e., $D = 0$, resulting in models 6Z and 15Z. 
If $D \ne 0$ is the cause of the 10\% dispersion,
then the apparent level of dispersion should change once
the eddy diffusivity is turned off. Figs. 9 and 10 also show that
even without any diffusion of the color field with respect to
the gas, the dispersion still appears to hover at a value
around 10\%, instead of approaching zero, as might have been
expected. These figures show that with $\alpha = 0.0001$,
the effects of diffusivity are minimal. However, in
models 6H and 15H, with $\alpha = 0.01$, the high diffusivity
leads to a high dispersion ($\sim$ unity), which does not appear
to be compatible with the isotopic evidence for some degree of
homogeneity in the nebula, and hence these models are not considered
further here.

 The results for models 6Z and 15Z imply that gravitational mixing 
is not 100\% efficient -- it is unable to completely homogenize an
initial spatial heteogeneity. This is probably because gravity is a
long-range force, able only to drive mixing over length scales
where significant gravitational forces occur. In the case of
these disk models, this corresponds to length scales comparable
to the spiral arms and clumps that develop, which tend to have
minimum length scales of no less than an AU or so, whereas the
radial extent of a single grid cell is 0.16 AU. Even the
convective-like motions that accompany the gravitationally-driven
large-scale motions do not appear to be able to homogenize
the color field down to the level of individual grid cells,
as these motions are themselves associated with the large-scale
motions being driven by the gravitational torques, which leads
to large-scale granularity at the 10\% level. 

 Finally, it is important to note that this 10\% level of spatial 
heterogeneity cannot be caused by the finite numerical viscosity
of the hydrodynamics code, because this intrinsic viscosity
affects both the gas density and the color field in exactly the same
way (their evolution equations are identical when $\alpha = 0$), 
so that numerical viscosity cannot produce variations in
the color field to gas density ratio.

\section{Discussion}

 The trend toward a steady state dispersion value consistent with
a low degree of spatial heterogeneity in these models implies that 
cosmochemical variations in inferred initial $^{26}$Al/$^{27}$Al 
ratios must be due primarily to temporal heterogeneity, preserving 
the role of SLRIs as accurate chronometers for the solar nebula.
While these models were originally intended to simulate the 
process of injection of SLRI onto the surface of the solar nebula, 
they are also roughly applicable to the situation regarding
the stable oxygen isotope anomalies, thought to be created
quasi-continuously at the disk surface at distances of $\sim$
30 AU (Lyons \& Young 2005). In a marginally gravitationally
unstable disk, the optically thin surface will be dynamically
evolving with a highly rippled surface (Jang-Condell \& Boss 2007).
One might expect then
that the right conditions for generation of the oxygen anomalies
will be a spatially and temporally varying process, rather
than a continuous process across the entire disk surface. If so,
then these models of instantaneous injection onto limited regions
of the disk surface should be applicable as well to the mixing
and transport of oxygen anomalies produced by UV photodissociation.

 Boss (2006) presented a scenario for the transport of $^{16}$O-poor
ice grains formed at the disk surface at 20 AU and beyond (Lyons \& Young
2005) inward to $\sim$ 2.5 AU on timescales of $\sim 10^3$, as indicated
by the present models and by those of Boss (2004a). This would change the 
oxygen composition of the inner disk from being $^{16}$O-rich to $^{16}$O-poor.
Krot et al. (2005a) and Zanda et al. (2006) suggested that the unaltered 
oxygen isotopic compositions of chondrules and refractory inclusions (RI) 
resulted from the RI forming first in an $^{16}$O-rich gas, followed by
chondrule formation after the nebular gas had evolved to a 
more $^{16}$O-poor composition. The present models support
this type of explanation for the oxygen isotope anomalies observed in
different components of primitive meteorites.

 The basic result that initially spatially heterogeneous
distributions tend to become homogenized to the extent that
the dispersion from the mean value hovers around $\sim$ 10\% 
appears to be the result of the physics of mixing and transport
in a marginally gravitationally unstable disk, rather than
a result of the statistics of sampling rare events, e.g., the
Poisson statistics of radioactive isotope decays. In the case of
Poisson statistics, if the average number of events recorded
in a given time interval is $N$, then the standard deviation 
of the number of events is $N^{1/2}$, which when divided by the 
average number of events is equal to $1/N^{1/2}$. This quantity
is equivalent to the dispersion defined in this paper, which
tends toward a value $\sim$ 10\%. If this value were
the result of Poission statistics, then, it would imply that
$N \sim 100$. It is unclear what the appropriate choice should
be for $N$, but one choice could be the number of grid cells
sampled. Given that the number of computational cells
in the disk midplane used to calculate the dispersion in
the present models is $\sim 25,000$, if one took $N = 25,000$,
and tried to use Poission statistics to predict a dispersion
based on this sampling of the finite number of grid cells,
the result would be $1/N^{1/2} \sim 0.006$, not 0.1. Hence
Poisson statistics cannot explain this result. In this context it
should be noted that radioactive decay of the SLRI are not included 
in the present calculations, as the evolutions cover times
($<$ 0.004 Myr) that are much shorter than the half-lives
of the SLRI (e.g., the half-life of $^{26}$Al is 0.7 Myr). 

\section{Conclusions}

 Refractory inclusions such as the CAIs have $^{26}$Al/$^{27}$Al ratios that 
scatter by about 10\% around the mean value of $\sim 4.5 \times 10^{-5}$ 
(MacPherson et al. 1995). This level of scatter in  $^{26}$Al/$^{27}$Al 
ratios appears to be consistent with the 10\% dispersion found here to 
follow from an apparently stable level of spatial heterogeneity in the 
nebula. Such a relatively low level of spatial heterogeneity implies 
that even larger variations in initial $^{26}$Al/$^{27}$Al ratios must 
be due primarily to temporal heterogeneity, preserving the crucial role 
of SLRIs like $^{26}$Al/$^{27}$Al as accurate chronometers for the solar 
nebula. This $\sim$ 10\% level of heterogeneity also appears to be
consistent with the preservation of the stable oxygen isotope anomalies 
produced by self-shielding at the outer disk surface (Lyons \& Young 2005),
followed by transport downward to the midplane and inward to regions of 
the nebula where such anomalies could also have been created but perhaps
not sustained because disk surface temperatures there were too high for 
stability of the water ice grains that carried the isotopic anomalies.

 We have shown that the degree of spatial heterogeneity in the solar
nebula required for the use of the $^{26}$Al/$^{27}$Al chronometer
and for the survival of the oxygen isotope anomalies is consistent
with the expectations for the mixing and transport of initially
highly spatially heterogeneous tracers in a marginally gravitationally 
unstable disk. Such a solar nebula appears to be required
for the formation of gas giant planets, such as Jupiter and Saturn, by
either the core accretion (Inaba et al. 2003) or disk instability 
(Boss et al. 2002) mechanisms. MRI is often thought to be the source 
of the turbulent viscosity that is assumed to drive disk evolution 
across large distances (e.g., Gail 2002, 2004; Ciesla \& Cuzzi 2006), 
but the presence of magnetically dead zones limits the applicability of
MRI-driven turbulence to magnetically live regions, such as the 
disk surfaces. Marginally gravitationally unstable disk models offer a 
means to self-consistently calculate the mixing and transport of tracers 
throughout the planet-forming region. This large-scale transport
seems to be required to explain observations of thermally processed,
crystalline silicates in comets and protoplanetary disks (Nuth, Rietmeijer, 
\& Hill 2002; van Boekel et al. 2005; Honda et al. 2006; Nuth \& 
Johnson 2006). Marginally gravitationally unstable disks have the
added attraction of providing a robust source of shock fronts
capable of thermally processing solids into the chondrules and 
other components found in the most primitive meteorites
(Desch \& Connolly 2002; Boss \& Durisen 2005).

\acknowledgments

 I thank Francis Albarede, James Lyons, and Joseph A. Nuth III 
for discussions, the referee for helpful comments, and Sandy Keiser for 
cluster and workstation management. This research was supported in 
part by the NASA Planetary Geology and Geophysics Program under grant 
NNG05GH30G and by the NASA Origins of Solar Systems Program under grant 
NNG05GI10G, and is contributed in part to the NASA Astrobiology Institute 
under grant NCC2-1056. Calculations were performed on the Carnegie Alpha 
Cluster, which was supported in part by NSF MRI grant AST-9976645.

\vfill\eject

\begin{figure}
\vspace{-2.0in}
\plotone{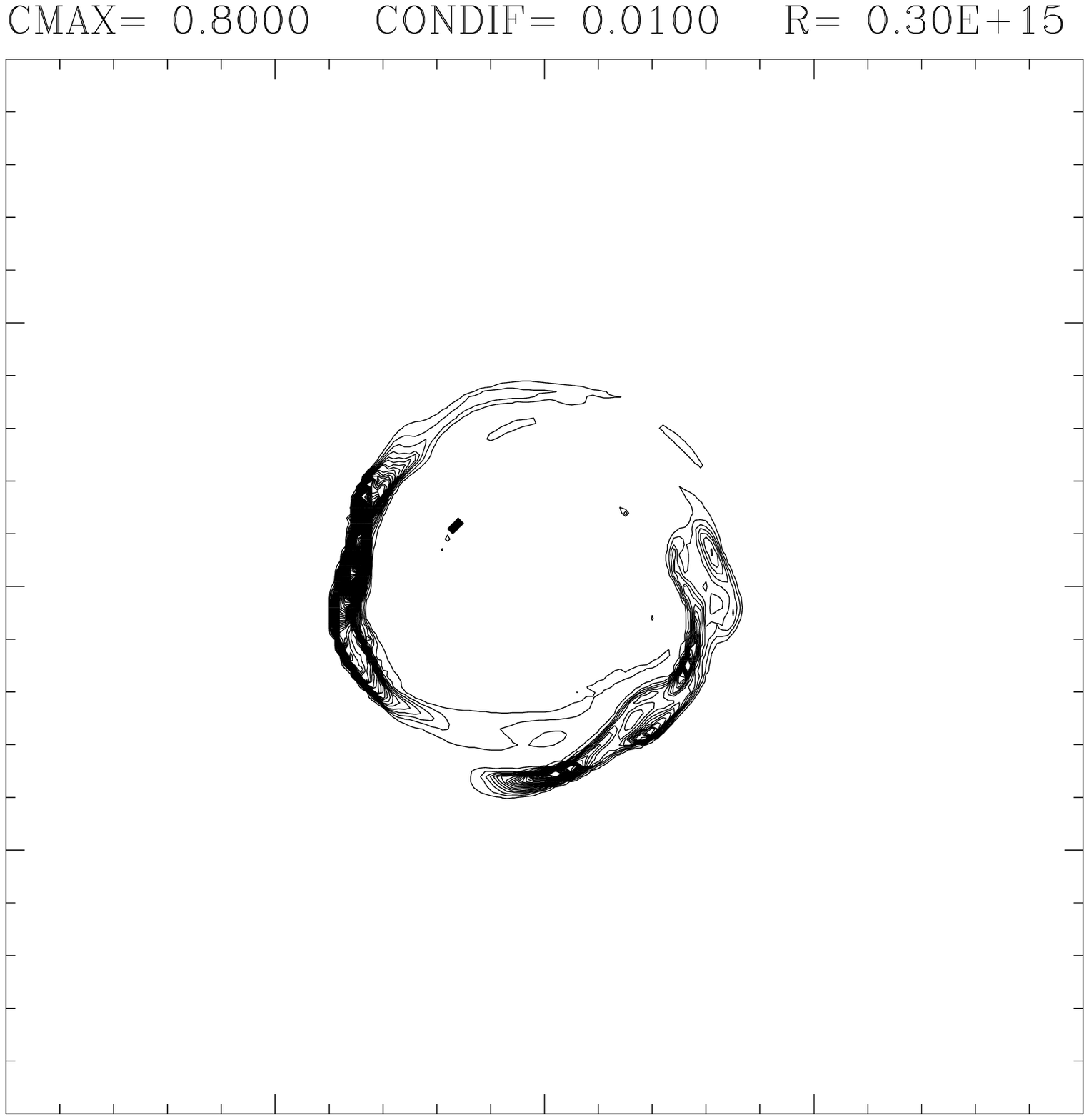}
\caption{Model 6L at 227 yr, showing linear contours of 
the color field density (e.g., number of atoms of $^{26}$Al cm$^{-3}$) 
in the disk midplane 27 yrs after the color field was sprayed onto 
the disk's surface in a 90 degree azimuthal sector between 5.5 and 
6.5 AU. Region shown is 20 AU in radius (R) with a 4 AU radius inner 
boundary. In this Figure, the contours represent changes in the color 
field density by 0.01 units (CONDIF) on a scale normalized by the 
initial color field density of 1.0, up to a maximum value of 0.8 (CMAX). 
At this time, the color field has reached the midplane and begun to 
spread radially inward and outward.}
\end{figure}

\begin{figure}
\vspace{-2.0in}
\plotone{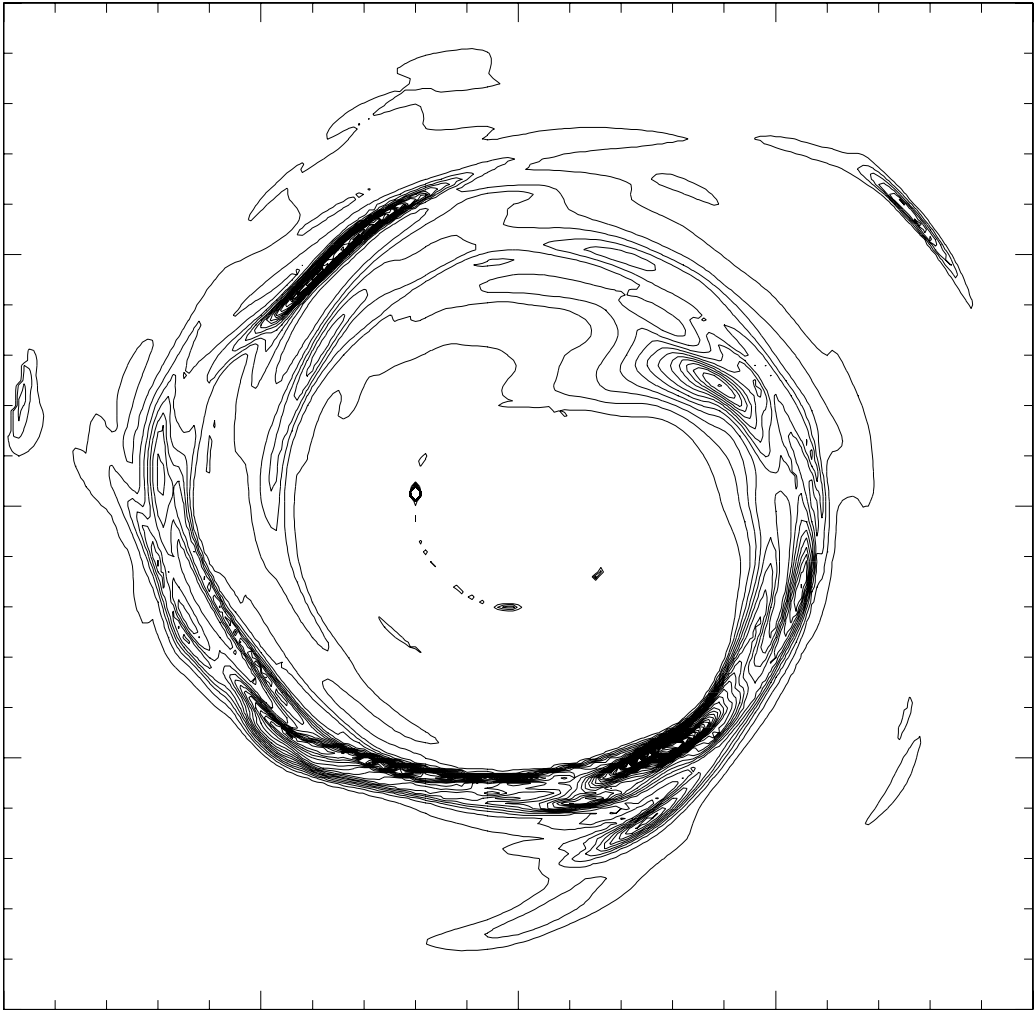}
\caption{Same as Fig. 1, but after 760 yr. While the color field has spread 
throughout the radial extent of the disk, the spatial distribution is 
highly heterogeneous, with the highest concentrations residing inside the 
spiral arms of the disk.}
\end{figure}

\begin{figure}
\vspace{-2.0in}
\plotone{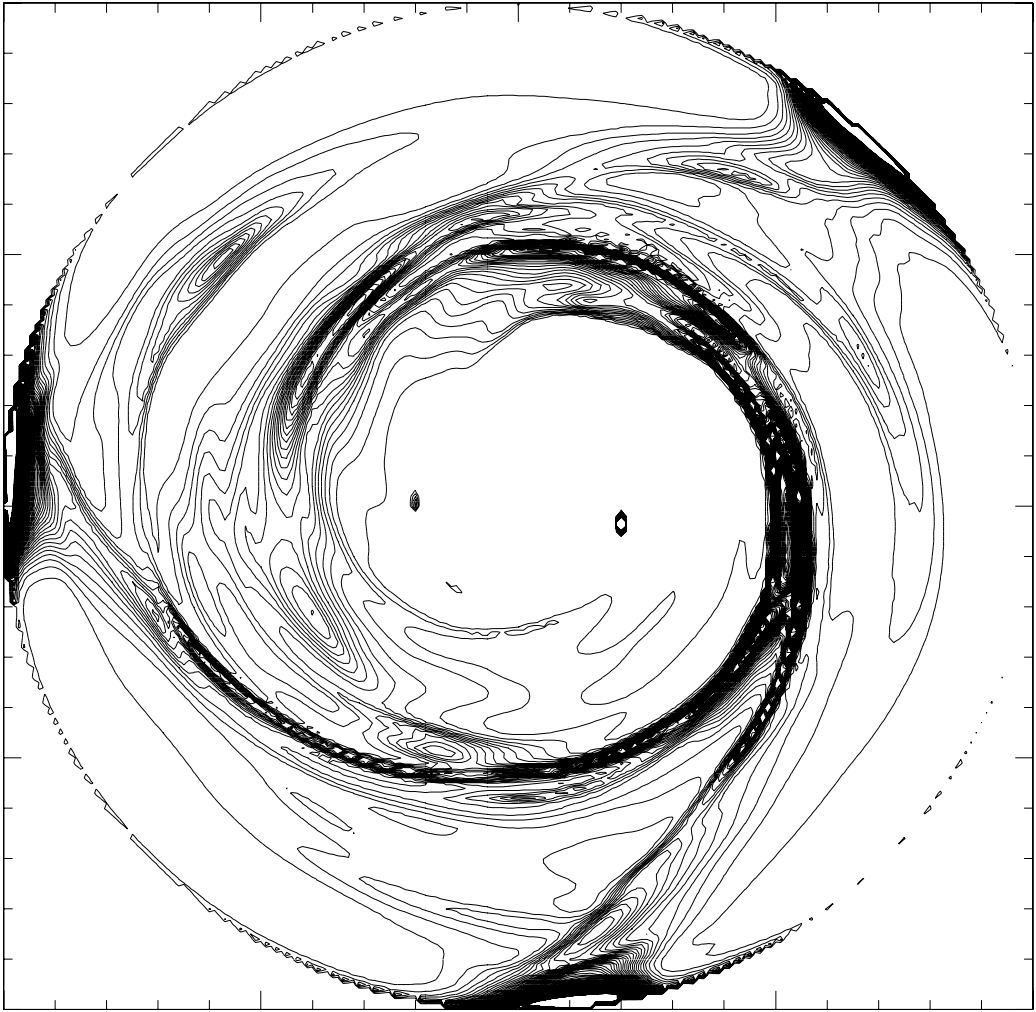}
\caption{Same as Fig. 1, but after 3484 yr. The color clumps on
the inner and outer boundaries are artifacts of the way the boundaries
are handled.}
\end{figure}

\begin{figure}
\vspace{-2.0in}
\plotone{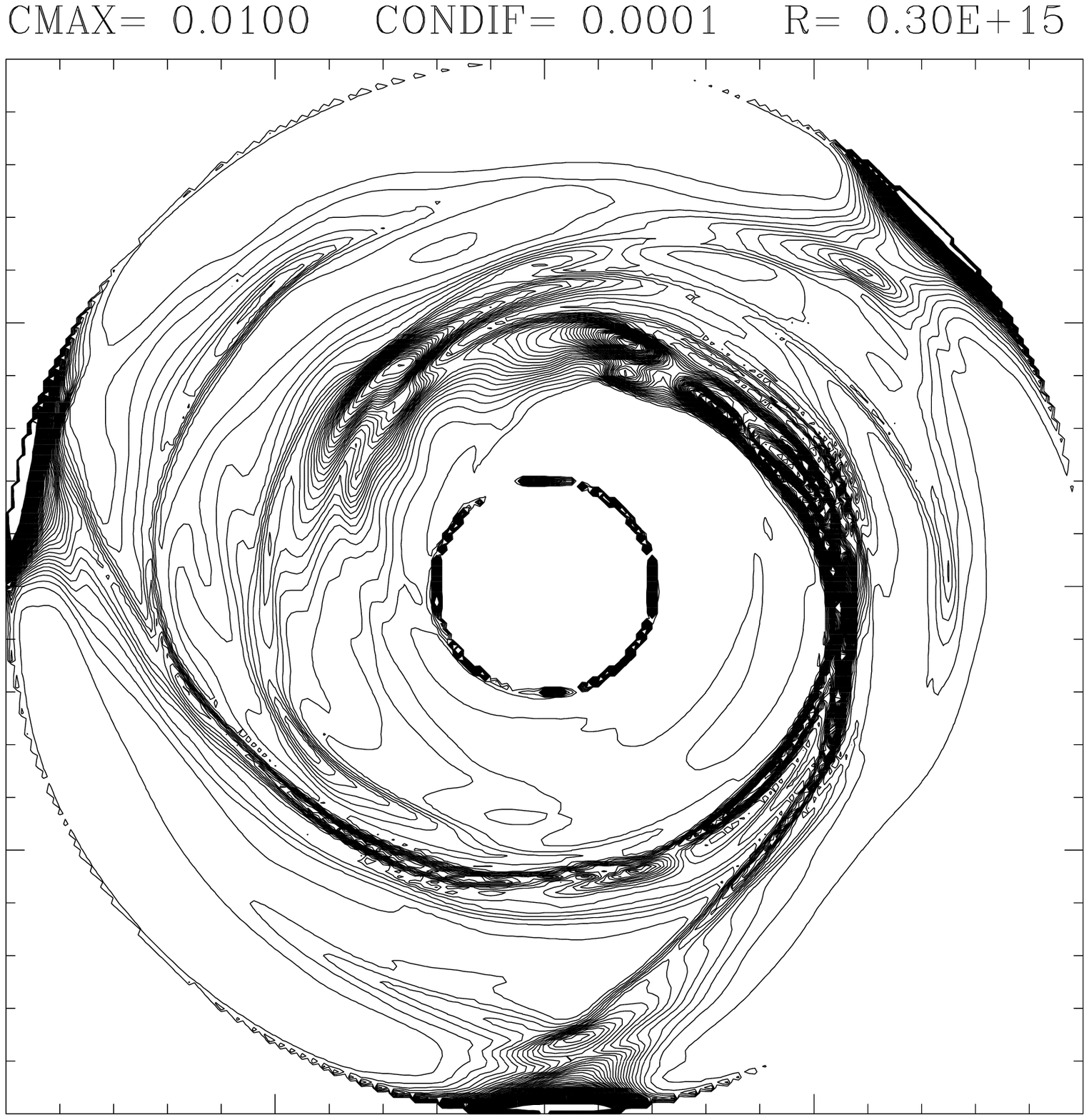}
\caption{Same as Fig. 1, but for model 6Z (no eddy diffusivity) at 
essentially the same time (3483 yr) as model 6L in Figure 3. The color 
fields are nearly identical for models 6Z and 6L.}
\end{figure}

\begin{figure}
\vspace{-2.0in}
\caption{Logarithmic contours of the color field density divided by 
the disk gas density (i.e., log of the abundance ratio $^{26}$Al/$^{27}$Al) 
for model 6L at a time of 3484 yr, plotted in the same manner as in Fig. 1.
Contours represent changes by factors of 1.26 up to a 
maximum value of 8.0, on a scale defined by the gas disk density.
The abundances of SLRI injected by
a supernova are rapidly homogenized to within $\sim$ 25\%, except
for inside very low gas density regions adjoining the artificial
inner and outer boundaries. [figure deleted to fit on astro-ph -- can
be seen in files on http://www.dtm.ciw.edu/boss/ftp/nebulaviii.]}
\end{figure}

\begin{figure}
\vspace{-2.0in}
\plotone{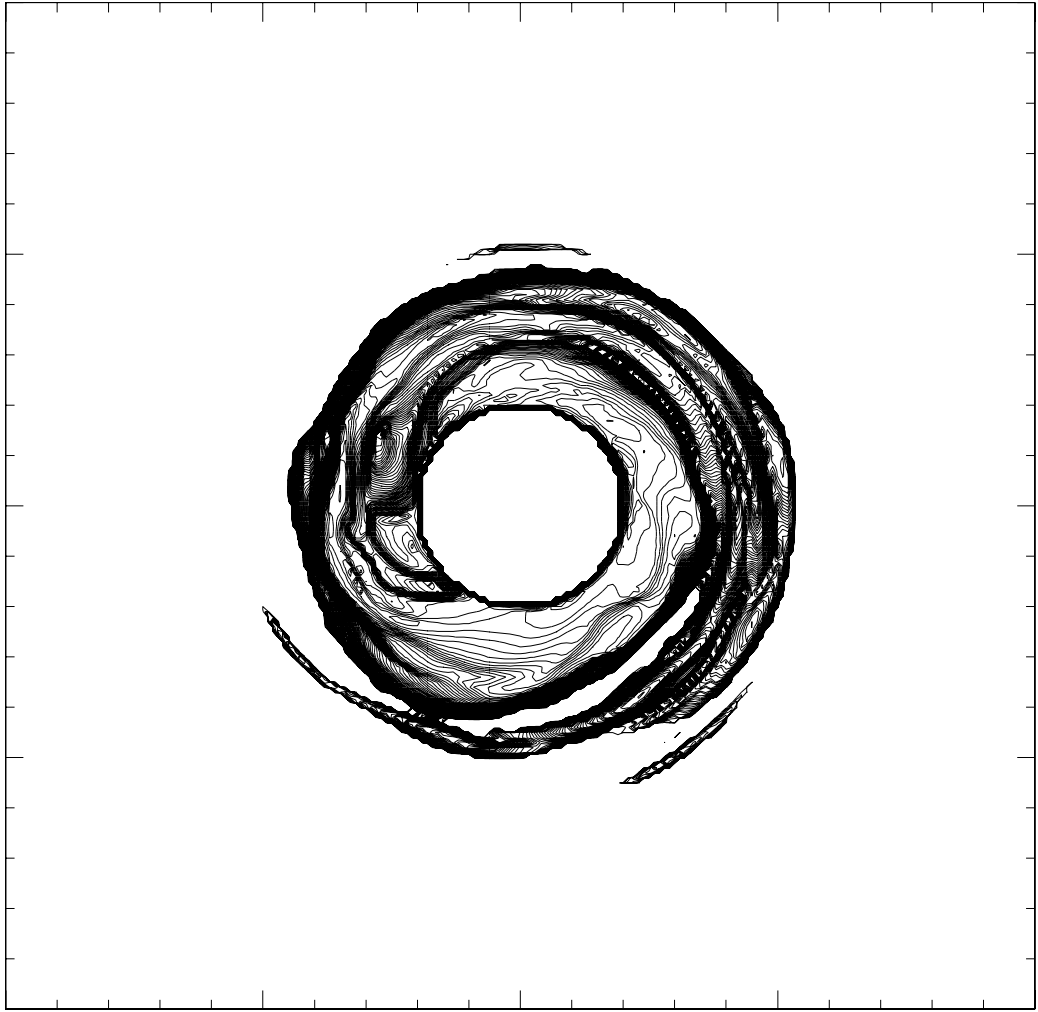}
\caption{Same as Fig. 5 for model 6L, but at an earlier time of
227 yr (same time as Fig. 1). The color/gas density ratio is
highly non-uniform at this early phase, before the marginally
gravitationally unstable disk has had a chance to transport the
color field radially inward and outward and to mix the field 
with the disk gas.}
\end{figure}

\begin{figure}
\vspace{-2.0in}
\plotone{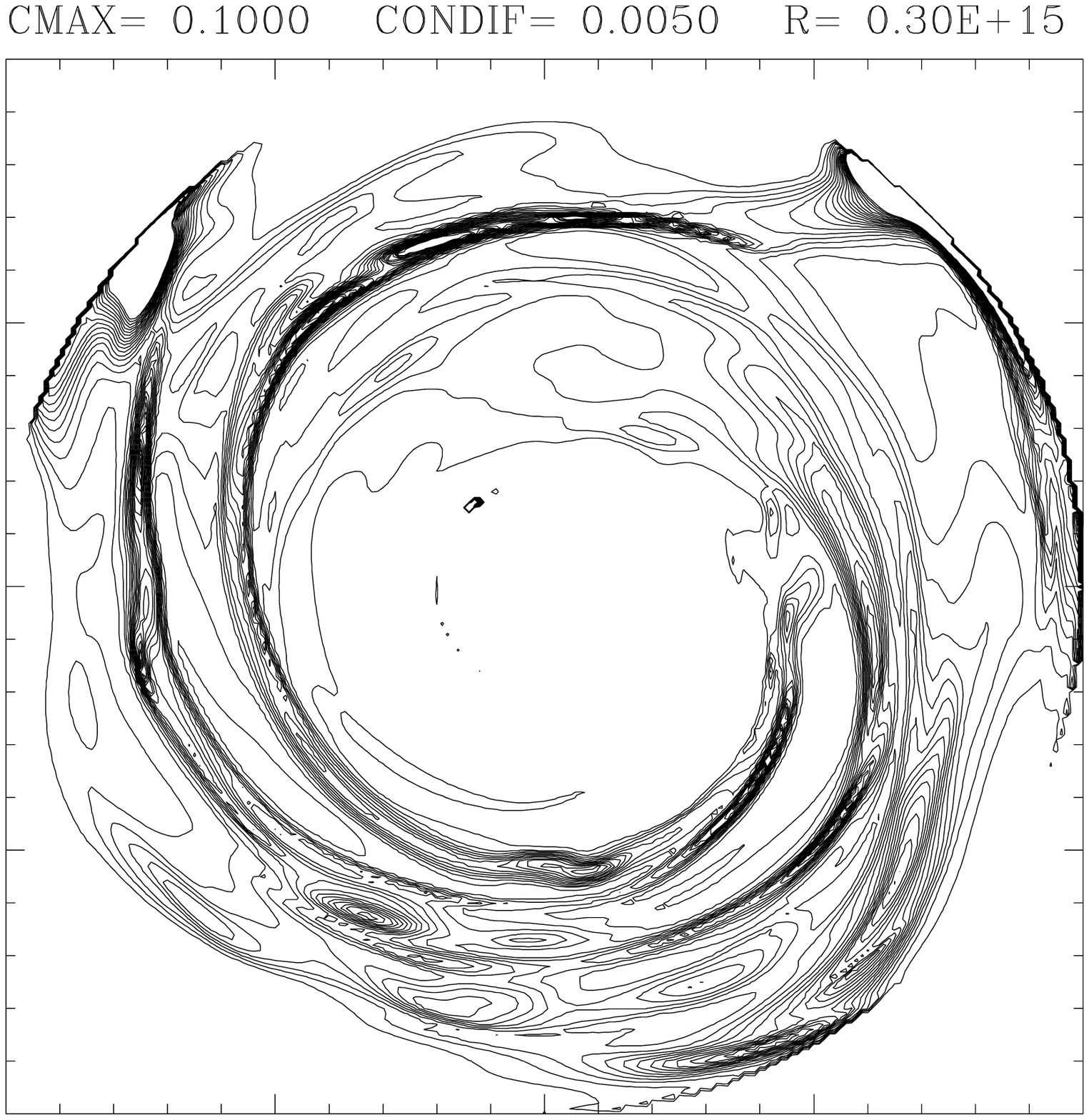}
\caption{Same as Fig. 1, but for model 15L at a time of 754 yr.
In this model, the color is injected onto the disk's surface
in a 90 degree azimuthal sector at an orbital radius of 15 AU 
instead of 6 AU.}
\end{figure}

\begin{figure}
\vspace{-2.0in}
\caption{Same as Fig. 5, but for model 15L at a time of 3491 yr.
Once again, the color/gas density ratio has become highly uniform
as a result of mixing and transport, with the exception of the
low density regions at the boundaries.
[figure deleted to fit on astro-ph -- can
be seen in files on http://www.dtm.ciw.edu/boss/ftp/nebulaviii.]}
\end{figure}

\begin{figure}
\vspace{-2.0in}
\plotone{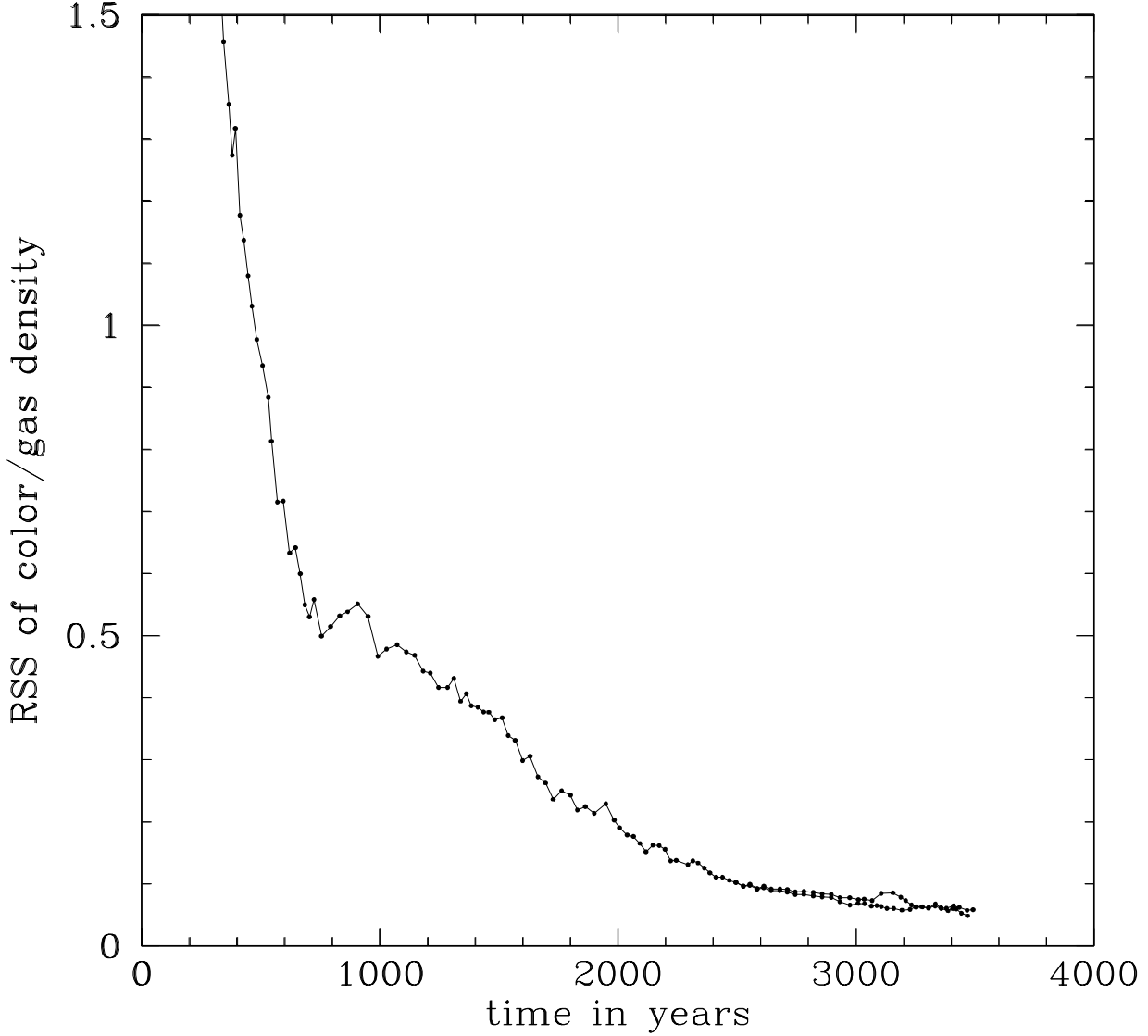}
\caption{Time evolution of the dispersion from the mean (i.e., standard
deviation, or the root of the sum of the squares [RSS]of the differences
from the mean) of the color field density divided by the gas density
(e.g., $^{26}$Al/$^{27}$Al abundance ratio) in the disk midplane in
model 6L. The color field is sprayed onto the disk surface at a 
time of 200 yrs. Starting from high values (RSS at 200 yrs is $>>$ 1),
the dispersion decreases on a timescale of $\sim 1000$ yrs, then
approaches a steady state value of $\sim$ 10\%. After $\sim$ 2600 yr,
the results for model 6Z are plotted as well, resulting in the 
slightly lower sequence of values.}
\end{figure}

\begin{figure}
\vspace{-2.0in}
\plotone{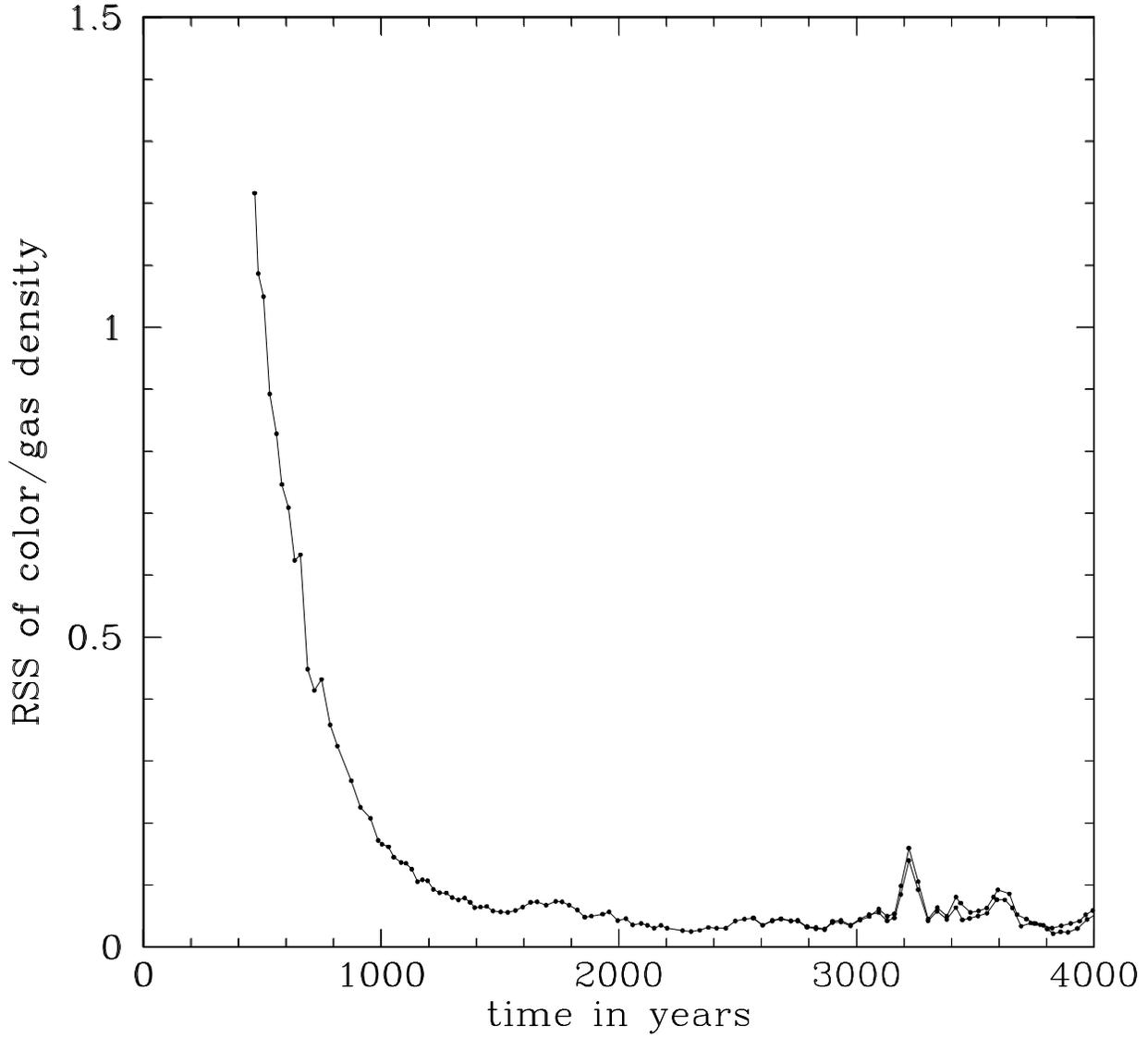}
\caption{Same as Fig. 9, but for models 15L and 15Z, except that
in this case the values for model 15Z are slightly higher than
those for 15L.}
\end{figure}

\end{document}